\documentstyle[preprint,aps]{revtex}
\tighten
\begin{document}
\draft
\title{Coulomb drag in mesoscopic rings}
\author{T. V. Shahbazyan and S. E. Ulloa}
\address{Department of Physics and Astronomy,
Condensed Matter and Surface Science Program, Ohio University,
Athens, OH 45701-2979}
\maketitle

\begin{abstract}
We develop a Luttinger liquid theory of the Coulomb drag of persistent
currents flowing in concentric mesoscopic rings, by incorporating
non-linear corrections to the electron dispersion relation. 
We demonstrate that at low temperatures, interactions between electrons
in different rings generate an additional phase and thus alter the 
{\em period} of Aharonov-Bohm oscillations. The resulting 
{\em nondissipative} drag depends strongly on the {\em relative parity} 
of the electron numbers. We also show that interactions set a new
temperature scale below which the linear response theory does not
apply at certain values of external flux.
\end{abstract}
\narrowtext

During the last decade, persistent currents (PC's) in mesoscopic rings
have attracted significant interest both theoretically
\cite{but83,che89,thmet,los92,yu92,mul93}
and experimentally\cite{lev90,cha91,mai93}.  
Much of this attention was due to a large discrepancy between
experimentally observed  current amplitudes in disordered metallic
rings\cite{lev90,cha91}, and theoretical predictions based on a 
single-particle picture\cite{che89}. Yet unresolved, this puzzle
has generated a number of theoretical works\cite{thmet} on the role of
electron-electron interactions in multi-channel disordered rings.

At the same time, in clean single-channel rings the
theory predicts that PC's at low temperatures should not be
affected by interactions\cite{leg91,los92,mul93}, and exhibit
Aharonov-Bohm oscillations  as a function of flux with
the same period  and amplitude as for non-interacting
electrons. Results of a recent experiment on a single 
semiconductor ring with low number of channels are in agreement
with these predictions\cite{mai93}. 

On the other hand, interactions should become essential in a system
consisting of a pair of clean 1D rings with different radii, placed
concentrically, as is shown in Fig.~1. If the rings were isolated,
PC's in each ring would oscillate with a period determined by its
radius. Inter-ring interactions will change the oscillation pattern by
causing {\em Coulomb drag} of PC's, as we show below.

There are, in general, two physical mechanisms for the current drag.
The first mechanism\cite{discdrag}, which originates from ``friction'' 
between two subsystems caused by scattering of carriers in one
subsystem  by density fluctuations in the other, has been widely
studied during recent years\cite{boldrag}, following
experimental observation of the Coulomb drag\cite{expdrag}. At low
temperatures, the resulting transresistance behaves as 
$T^2$ ($T^2\ln T$ in a disordered system) and vanishes then with
decreasing $T$ as the phase space for scattering shrinks. The second,
nondissipative mechanism was pointed out by Rojo and
Mahan\cite{rojo92}, and is based on the observation that the 
van der Waals interaction between two current carrying subsystems is
modified, resulting in a finite current drag at zero temperature. 

The latter mechanism is, in fact, responsible for the PC drag in
mesoscopic rings. Inter-ring interactions change the ground state
energy of the system. As a result, PC in each ring, being a 
{\em thermodynamic} quantity, acquires a dependence on the flux 
penetrating the other ring. With decreasing temperature, this 
dependence should become sharper because of a singular (``sawtooth'')
shape of the PC as a function of flux at zero temperature. In the
vicinity of a ``tooth'', even a small perturbation may affect strongly
the {\em amplitude} of the PC. For this reason, an adequate theory of
the PC drag must be valid beyond linear response. 

In this paper we develop such a theory using the Luttinger liquid
(LL) approach to PC in 1D rings. This approach is based on Haldane's
concept of topological excitations in finite-size 1D
systems\cite{hal81}, and was recently extended by Loss\cite{los92} to
account for parity effects\cite{leg91} in the presence of external
flux. However, in its standard formulation, the LL approach does
not allow for current drag, which arises from the asymmetry between
electrons and holes. With electron dispersion linearized, the
asymmetry is lost and the electron and hole drags completely
compensate each other. In LL formalism, the electron-hole symmetry
manifests itself in a complete separation between zero modes, which
carry PC, and bosonic fields, which are responsible for inter-ring
interactions. We derive an interaction-induced correction to the zero
mode spectrum by incorporating the lowest order non-linear correction
to the electron dispersion relation. Further, we show that at low
temperatures, the entire effect of inter-ring interactions is to
generate a phase which, in turn, leads to PC drag by changing the flux
seen by the electrons.  

We begin by recalling the LL approach to interacting (spinless)
electrons in a 1D ring with circumference $L_i$ ($i=1,2$) threaded by 
magnetic flux $\phi_i$\cite{hal81,los92}. By linearizing the spectrum
near the Fermi points, the electrons are decomposed into right and
left moving fermions with $\psi_{i\alpha}(x)$ ($\alpha=\pm$)
satisfying twisted boundary conditions
$\psi_{i\alpha}(x+L_i)=
(-1)^{N_i} e^{2\pi\phi_i/\phi_0}\psi_{i\alpha}(x)$, where the sign
factor reflects the dependence of the ground state on the parity
of the particle number $N_i$ \cite{leg91,los92} and $\phi_0$ is the
flux quantum. The fermion operator can be expressed through bosonic
fields as 
$\psi_{i\alpha}(x)=(2\epsilon L_i)^{-1/2}e^{-i\varphi_{i\alpha}(x)}$,
with $\varphi_{i\alpha}(x)=[\theta_i(x)-\alpha\varphi_i(x)]/2$. The 
fields $\varphi_i(x)$ and $\theta_i(x)$ can be presented
as\cite{remark1} 

\begin{eqnarray}\label{bosefield}
\varphi_i(x)&=&\varphi_{iJ}+M_i2\pi x/L_i+\overline{\varphi}_i(x),\\
\theta_i(x)&=&\theta_{iM}+(J_i-2\Phi_i)2\pi (x+L_i/2)/L_i+
\overline{\theta}_i(x),
\end{eqnarray}
where $M_i$ and $J_i$ are fermion number and (topological) current
operators, and $\theta_{iM}$ and $\varphi_{iJ}$ are their conjugates: 
$[\theta_{iM},M_{i'}]=[\varphi_{iJ},J_{i'}]=2i\delta_{ii'}$. For
further convenience, we have incorporated both the flux and parity
dependence into a single quantity $\Phi_i$, which is defined as
$\Phi_i=\phi_i/\phi_0$ for $N_i$ odd, and $\Phi_i=\phi_i/\phi_0-1/2$ 
for $N_i$ even. Owing to the boundary condition, the eigenvalues of
$J_i$ and $M_i$ satisfy the selection rule that their sum is even.
The periodic fields $\overline{\varphi}_i(x)$ and
$\overline{\theta}_i(x)$ have the usual representation in terms of
bosonic creation and annihilation operators

\begin{eqnarray}\label{perfield}
\overline{\varphi}_i(x)&=&
\sum_{k_i\neq 0}\Biggl|{2\pi\over L_i k_i}\Biggr|^{1/2}
e^{ik_ix}(a_{ik_i}^{\dag}+a_{ik_i}),\\
\overline{\theta}_i(x)&=&
\sum_{k_i\neq 0}\Biggl|{2\pi\over L_i k_i}\Biggr|^{1/2}
\mbox{sgn}(k_i)e^{ik_ix}(a_{ik_i}^{\dag}-a_{ik_i}),
\end{eqnarray}
where $k_i=2\pi p/L_i$ ($p$ being an integer), and the
regularization factor $\exp(-\epsilon |k_i|L_i/2\pi)$ is implicit.
In the absence of backscattering, $\varphi_i$ is related to the charge
density (relative to the background density $N_i/L_i$) by
$\rho_i=\partial_x\varphi_i/2\pi$. The eigenvalues of $M_i$ represent 
the numbers of extra electrons added to the ring while that of $J_i$
are the excess of left over right moving fermions\cite{hal81}.

In terms of the bosonic fields $\varphi_i$, the (normal ordered)
Hamiltonian of free fermions with linear dispersion reads

\begin{eqnarray}\label{freeham}
H_0=\sum_i {v_i\over 8\pi}\int_0^{L_i}dx
:(4\pi\Pi_i)^2+(\partial_x\varphi_i)^2:\, ,
\end{eqnarray}
where $v_i$ is the Fermi velocity and  
$\Pi_i\equiv P_i-\Phi_i/L_i=\partial_x\theta_i/4\pi$, $P_i$ being 
the canonical momentum.

In calculating PC's it is convenient to go to the
Lagrangian formulation and present the partition function $Z$ in
terms of a functional integral. The free (Euclidian) action has the
form 

\begin{eqnarray}\label{freeaction}
S_0=\sum_i \int_0^{\beta}d\tau\int_0^{L_i}dx \Biggl\{{1\over 8\pi v_i}
[(\partial_{\tau}\varphi_i)^2+v_i^2(\partial_x\varphi_i)^2]
-{i\over L_i}(\Phi_i+\delta_{i})\partial_{\tau}\varphi_i\Biggr\},
\end{eqnarray}
where $\beta$ is the inverse temperature and $\varphi(\tau,x)$
has a decomposition (up to a constant)

\begin{eqnarray}\label{decomp}
\varphi_i(\tau,x)=2\pi n_i\tau/\beta
+2\pi m_ix/L_i+\overline{\varphi}_i(\tau,x).
\end{eqnarray}
Here zero modes $n_i$ and $m_i$ are winding numbers in $\tau$ and $x$
directions, respectively, and $\delta_{i}$ in (\ref{freeaction})
enforces the selection rule by taking values $\delta_{i}=0~(1/2)$ for
$m_i$ even (odd)\cite{los92}. Note that $m_i$ coincides with the 
eigenvalue of $M_i$. In the following we assume the number of
electrons in the rings fixed and restrict ourselves to the $m_i=0$
(with $\delta_i=0$) sector.

The partition function is given by 
$Z=\int D\varphi e^{-S}$, where $S=S_0+S_{int}$. Here the term

\begin{eqnarray}\label{intaction}
S_{int}={1\over 8\pi^2}\sum_{ij}\int_0^{\beta}d\tau
\int_0^{L_i}dx\int_0^{L_j}dx'
\partial_x\varphi_iV_{ij}(x,x')\partial_{x'}\varphi_j,
\end{eqnarray}
with $V_{ij}$ being the Coulomb potential, describes intra- and
inter-ring interactions, and the measure $D\varphi$ includes the sum
over zero modes together with the integral over periodic fields
$\overline{\varphi}_i$. The PC is found by separating out the
contribution from the zero-modes. With $\varphi_i(\tau,x)$ of the form
(\ref{decomp}) (with $m_i=0$), the latter are completely decoupled and
the action takes the form 

\begin{eqnarray}\label{action}
S=\sum_i \Biggl({n_i^2 \over \beta T_{i}}
-2\pi in_i\Phi_i\Biggr)+\overline{S},
\end{eqnarray}
where
$\overline{S}=S_0[\overline{\varphi}]+S_{int}[\overline{\varphi}]$ is
flux-independent and $T_{i}=2v_i/\pi L_i$ is the temperature
scale set by the finite size of each ring. Thus, the PC, 
${\cal J}_i=\beta^{-1}\partial_{\phi_i}\ln Z$, is carried by zero
modes whose spectrum is unaltered by inter-ring interactions. This 
is a consequence of the linearization of the electron
dispersion and the resulting quadratic form of the bosonic action
$S_0$. Physically, linearization lifts the electron-hole asymmetry,
which is responsible for the current drag, so that PC
in each ring is sustained by its own flux $\phi_i$.

Let us now consider the lowest non-linear correction to the fermion
dispersion relation. Following \cite{hal81}, we take the correction to
the linearized fermion Hamiltonian as

\begin{eqnarray}\label{nlham}
H_{nl}=-{1\over 2m_e}\sum_{i\alpha}\int_0^{L_i}dx
:\psi_{i\alpha}^{\dag}(x)\partial_x^2\psi_{i\alpha}(x):\, ,
\end{eqnarray}
where $m_e$ is electron mass. Transforming (\ref{nlham}) with the help
of bosonic representation of $\psi_{i\alpha}(x)$, we obtain 

\begin{eqnarray}\label{nlhambos}
H_{nl}={1\over 48\pi m_e}\sum_i \int_0^{L_i}dx
:(\partial_x\varphi_i)^3+3\partial_x\varphi_i(4\pi\Pi_i)^2:\, ,
\end{eqnarray}
which means that the corresponding bosonic theory is no longer
free. Regarding $H_{nl}$ as a small term, we can obtain a correction
to the action by adding (\ref{nlhambos}) to (\ref{freeham}) and
repeating the steps leading to the functional-integral representation
of $Z$\cite{remark2}. To first order in $1/m$, the correction takes
the form

\begin{eqnarray}\label{nlaction}
S_{nl}={1\over 48\pi m_e}\sum_i \int_0^{\beta}d\tau\int_0^{L_i}dx 
\Biggl[(\partial_x\varphi_i)^3-
{3\over v_i^2}\partial_x\varphi_i(\partial_{\tau}\varphi_i)^2
\Biggr].
\end{eqnarray}
Using (\ref{decomp}) (with $m_i=0$), we separate out the
zero-mode contribution in $S_{nl}$ by writing

\begin{eqnarray}\label{nlaction2}
S_{nl}=\sum_i {n_i\over 4\beta m_ev_i^2}
\int_0^{\beta}d\tau\int_0^{L_i}dx\,  
\overline{\varphi}_i\partial_{\tau}\partial_x\overline{\varphi}_i
+\overline{S}_{nl},
\end{eqnarray}
where $\overline{S}_{nl}$ depends only on fields
$\overline{\varphi}_i$, which are now coupled to the zero modes via
the first term.  Adding (\ref{nlaction2}) to (\ref{action}), we then
perform the functional integral 
$e^{-\beta\overline{F}}=\int D\overline{\varphi}
e^{-\overline{S}-S_{nl}}$. 
Note that after the zero modes are separated out, the small term
$\overline{S}_{nl}$ can be neglected and the remaining gaussian  
integral iexplicitly evaluated. The resulting $\overline{F}(n_1,n_2)$
admits an expansion in terms of $1/m_e$ with the odd orders vanishing
due to translational invariance. Thus, to the first non-vanishing order 

\begin{eqnarray}\label{fbos}
\overline{F}(n_1,n_2)=\overline{F}_0+
\sum_{ij}\overline{F}_{ij}n_in_j,
\end{eqnarray}
where $\overline{F}_0$ is $n_i$ and $\phi_i$ independent. The second
term in (\ref{fbos}), which is to be combined with the first term in
(\ref{action}), represents the correction sought to the zero-mode
spectrum. The diagonal coefficients $F_{ii}$ can be absorbed into
$T_{i}$ via renormalization of the Fermi velocities and 
do not play any role in the following\cite{remark3}. Factorizing out
the zero-mode part of the partition function, 
$Z_0\equiv e^{-\beta F_0}=e^{\beta\overline{F}_0}Z$, we finally arrive
at 

\begin{eqnarray}\label{part}
Z_0=\sum_{n_1n_2}\exp\Biggl(-{1\over\beta}\sum_{ij}a_{ij}n_in_j
+2\pi i\sum_i n_i\Phi_i\Biggr),
\end{eqnarray}
where $a_{ii}=T_{i}^{-1}$, and the parameter 
$a_{12}=\beta^2\overline{F}_{12}$, given by 

\begin{eqnarray}\label{a12}
a_{12}=
{-1\over 16 m_e^2 v_1^2v_2^2\beta}\int dz_1 \int dz_2 
\partial_{\tau_1}\partial_{x_1}\overline{D}_{12}(z_1,z_2)
\partial_{\tau_2}\partial_{x_2}\overline{D}_{21}(z_2,z_1),
\end{eqnarray}
describes {\em the coupling between zero modes in different rings}. 
Here $z_i=(\tau_i,x_i)$ and $\overline{D}_{12}(z_1,z_2)$ is
the nondiagonal part of the boson Green function,
$\overline{D}_{ij}(z_1,z_2)=
\langle \overline{\varphi}_i(z_1)\overline{\varphi}_j(z_2)\rangle$,
calculated from the action $\overline{S}$. Due to the azimuthal
symmetry of the system, a convenient basis for $\overline{D}_{ij}$ is
given by angular-momentum eigenfunctions, in which the
Fourier--transform of the inverse Green function reads

\begin{eqnarray}\label{Dij}
\overline{D}_{ij}^{-1}(\omega,p)={\delta_{ij}\over 4\pi v_i L_i}
\Biggl[\omega^2+v_i^2\Biggl({2\pi p\over L_i}\Biggr)^2\Biggr]+
{p^2\over L_iL_j}V_{ij}(p),
\end{eqnarray}
where $V_{ij}(p)$ is the Fourier-transform of the Coulomb
interaction. For $d\ll L_i$ the interaction is given by

\begin{eqnarray}\label{Vij}
V_{ij}(p)\simeq 
{2e^2\over\kappa (L_iL_j)^{1/2}}
K_0\Biggl[{2\pi (p+1/2) d_{ij}\over (L_iL_j)^{1/2}}\Biggr],
\end{eqnarray}
where $\kappa$ is the dielectric constant, $K_0$ is the modified
Bessel function, and we used the notation $d_{ij}=d$ for $i\neq j$,
and  $d_{ii}=w$, $w$ being the width of the rings. The expression for
$a_{12}$ then takes the form

\begin{eqnarray}\label{a12four}
a_{12}=
{-\pi^2\over 4 m_e^2 v_1^2v_2^2 L_1^2 L_2^2\beta}
\sum_{\omega p}\omega^2p^2[\overline{D}_{12}(\omega,p)]^2.
\end{eqnarray}
The frequency sum in (\ref{a12four}) is straightforward and yields

\begin{eqnarray}\label{a12freq}
a_{12}=
{-2\pi^2\over m_e^2 v_1v_2 L_1 L_2 (T_{1}T_{2})^{1/2}}
\sum_{p>0}pv_{12}{\partial\over\partial v_{12}}
\Biggl[ \Biggl({Q_{+}\over t_{+}}-{Q_{-}\over t_{-}}\Biggr)
{1\over Q_{+}^2-Q_{-}^2}\Biggl],
\end{eqnarray}
where

\begin{eqnarray}\label{defs}
t_{\pm}&=&\tanh\Biggl[{\pi^2\over 2}
\Biggl({T_{1}\over T}{T_{2}\over T}\Biggr)^{1/2}pQ_{\pm}\Biggr],~~~~
Q_{\pm}(p)=\sqrt{A(p)\pm B(p)},\\
A(p)&=&[r(1+v_{11})+r^{-1}(1+v_{22})]/2, ~~~~
B(p)=\sqrt{[r(1+v_{11})-r^{-1}(1+v_{22})]^2/4+v_{12}^2},\\
v_{ij}(p)&=&{2\over \pi^2}(T_{i}T_{j})^{-1/2}V_{ij}(p)
={2e^2\over\kappa \pi(v_iv_j)^{1/2}}
K_0\Biggl[{2\pi (p+1/2) d_{ij}\over (L_iL_j)^{1/2}}\Biggr],
\end{eqnarray}
and $r=(T_{1}/T_{2})^{1/2}$ characterizes the asymmetry between rings.
Let us obtain an estimate of $a_{12}$ for
identical rings, with $v_i=v$, $T_{i}=T_{0}$, and $L_i=L$. The
analysis of (\ref{a12freq}) shows that for  
$T\ll T_d\equiv T_0(L/d)\sqrt{1+\alpha\ln(2d/w)}$, where
$\alpha=2e^2/\kappa\pi v$ is dimensionless interaction constant, the
coupling $a_{12}$ is temperature {\em independent}. Since for $d\ll L$
one has $T_d\gg T_0$, and $T_0$ is the crossover temperature above
which the PC is exponentially suppressed, $a_{12}$ can
be regarded as a constant in the entire range of relevant
temperatures. For $d\gg w$ the screening is strong and $a_{12}$ is 
determined by the  first order in the inter-ring interaction. The
rhs of (\ref{a12freq}) then gives 

\begin{eqnarray}\label{a12est}
a_{12}\simeq -
{\alpha^2\over 32 T_0 d^2k_F^2[1+\alpha\ln(2d/w)]^{5/2}},
\end{eqnarray}
where $k_F$ if the Fermi momentum.
A $d^{-2}$ dependence (without logarithmic factor) was obtained
previously for infinite wires\cite{rojo92}.

Turning to the partition function (\ref{part}), we observe that since
the zero modes in different rings are now coupled, the PC in, say,
ring 1, ${\cal J}_1=-\partial F_0/\partial\phi_1$,  depends also on
the flux through ring 2. Since the coupling $a_{12}$ is small, one could
try to obtain a correction to the current by expanding the free energy to
first order in $a_{12}$. It is important to realize, however, that
at low temperatures such an expansion does {\em not} exist for all
values of $\Phi_i$. In order to make this point clear, let us rewrite
(\ref{part}) in a different form using Poisson's formula (omitting
constant prefactor) 

\begin{eqnarray}\label{part2}
Z_0=
\sum_{p_1p_2}\exp\Bigl[-\beta\sum_{ij}
(p_i-\Phi_i)c_{ij}(p_j-\Phi_j)\Bigr],
\end{eqnarray}
where $\hat{c}=\pi^2\hat{a}^{-1}$. For $a_{12}\ll a_{ii}$ one has for
diagonal elements $c_{ii}=\pi^2 T_{i}$ which, in the absence of
interactions, is the level spacing at the Fermi level. With such 
form of $Z_0$, the PC is given by

\begin{eqnarray}\label{curr}
{\cal J}_1(\Phi_1,\Phi_2)={2\over \phi_0}
\Biggl[ c_{11}(\langle p_1\rangle-\Phi_1)+
c_{12}(\langle p_2\rangle-\Phi_2)\Biggr],
\end{eqnarray}
where $\langle p_i\rangle$ stands for the average of $p_i$ calculated
from the partition function (\ref{part2}). Note that $\Phi_i$ in
(\ref{curr}) takes values in the interval $0<\Phi_i<1$, and 
${\cal J}_i$ is periodically continued outside of this interval. 
For further analysis, it is convenient to obtain an equation for 
$\langle p_i\rangle$. This can be done by making use of the following
identity

\begin{eqnarray}\label{ident}
\langle p_i\rangle=
{1\over 2}-{1\over 2}\sum_{n=-\infty}^{\infty}
\Biggl\langle\tanh\beta\Biggl[\sum_{j}c_{ij}(p_j-\Phi_j)
+c_{ii}(n+1/2-p_i)\Biggr]\Biggr\rangle,
\end{eqnarray}
which can be readily derived by substituting 
Jacoby's product formula for the $\theta_{3}$-function\cite{abram} in
place of the sum over variable $p_i$ in the double sum (\ref{part2})
(one notices, for example, that for $c_{12}=0$ the rhs of
(\ref{ident}) is $p$-independent and $\langle p_i\rangle-1$ reduces
to the log-derivative of the $\theta_{3}$-function). For low
temperatures, 
$T\ll c_{ii}$, the fluctuations of $p_i$ are suppressed and all 
moments factorize, $\langle p_i^m\rangle=\langle p_i\rangle^m$, which
allows one to replace $p_i$ by $\langle p_i\rangle$ in the rhs of
(\ref{ident}). It is also easy to see that for such temperatures, all
terms in the sum with $n\neq 0$ cancel each other out (up to
exponentially small corrections). Then (\ref{ident}) simplifies to the
system  

\begin{mathletters}\label{syst}
\begin{eqnarray}
\langle p_1\rangle=
f_0\Bigl[c_{11}(1-2\Phi_1)+2c_{12}(\langle p_2\rangle-\Phi_2)\Bigr],\\
\langle p_2\rangle=
f_0\Bigl[c_{22}(1-2\Phi_2)+2c_{12}(\langle p_1\rangle-\Phi_1)\Bigr],
\end{eqnarray}
\end{mathletters}
where $f_0(x)=(e^{\beta x}+1)^{-1}$ is the Fermi function. This
system, together with (\ref{curr}) determines the PC at low
temperatures, $T\ll c_{ii}=\pi^2T_{i}$. In the absence of inter-ring
coupling ($c_{12}=0$) one recovers the PC for an isolated ring, 

\begin{eqnarray}\label{J0}
{\cal J}_{i}^0(\Phi_i)=2I_{i}\Biggl\{
f_0\Bigl[c_{ii}(1-2\Phi_i)\Bigr]-\Phi_i
\Biggr\},
\end{eqnarray}
where $I_{i}=c_{ii}/\phi_0=ev_i/L_i$ is the current amplitude. The
current ${\cal J}_{i}^0$ is a linear function of flux except in the 
interval $\Phi_i-1/2\sim T/c_{ii}\ll 1$ in which it rapidly changes from
$-I_{i}$ to $I_{i}$. Note that $\Phi_i$ depends on the parity of the
total number of electrons, $N_i$, resulting in diamagnetic
(paramagnetic) current for $N_i$ odd (even)\cite{leg91,los92}.

Turning the coupling on, PC's can be found by iterating the system
(\ref{syst}). First, in the argument of the Fermi function we
substitute  $\langle p_i\rangle$ expressed via  ${\cal J}_i$ by
neglecting the second term in (\ref{curr}) (and in a similar
expression for ${\cal J}_2$). Substituting (\ref{syst}) back into
(\ref{curr}), we finally obtain 

\begin{eqnarray}\label{J}
{\cal J}_1(\Phi_1,\Phi_2)={\cal J}_{1}^0(\Phi_1-\delta_1),~~
\delta_1=b{\cal J}_{2}(\Phi_1,\Phi_2),
\end{eqnarray}
with $b=\phi_0 c_{12}/2c_{11}c_{22}=-\phi_0a_{12}/2\pi^2$.
Eq.~(\ref{J}), which is our main result, describes the mutual
dependence (Coulomb drag) of PC's in coupled rings. It shows that
electrons encircling the ring $i$ acquire additional (Berry's) phase 
proportional to PC in the ring $j$. Thus, {\em inter--ring
interactions change the period of the Aharonov-Bohm oscillations}.
In particular, peak positions of PC get shifted by an amount 
$\delta_i$. Remarkably, the resulting oscillation pattern depends
strongly on the {\em relative parity} of electron numbers $N_i$, as it
can be seen in Fig.~2. We emphasize that in the vicinity of the peak, 
$\Phi_i\simeq 1/2$, PC (\ref{J}) {\em cannot} be expanded in
$\delta_i$ if the temperature is low enough:  
$T\lesssim T^{\ast}\equiv c_{12}$. Similarly, 
${\cal J}_{j}(\Phi_1,\Phi_2)$ in the expression for $\delta_i$ cannot
be replaced by ${\cal J}_{j}^0(\Phi_j)$ for $\Phi_j\simeq 1/2$. 
If $\Phi_i$ is not too close to $1/2$, one can neglect $\delta_i$ in
the argument of the Fermi function and the correction to PC takes the
form $\delta{\cal J}_{i}/I_i=2b{\cal J}_{j}$  (note that $b$ is
positive). 

Eq.~(\ref{J}) was obtained for temperatures lower than $T_i$. With
increasing $T$, the PC amplitude gets damped due to the loss of
phase coherence. Expanding the exponent in (\ref{part2}) in terms of
$c_{12}$, the correction to the free energy  can be written
as

\begin{eqnarray}\label{deltaf}
\delta F_0=\phi_0b{\cal J}_{1}^0(\Phi_1){\cal J}_{2}^0(\Phi_2),
\end{eqnarray}
with ${\cal J}_{i}^0(\Phi_i)
=(4I_i/\pi)(T/T_i)e^{-T/T_i}\sin(2\pi\Phi_i)$\cite{los92}.
Correspondingly, the correction to PC for $T\gg T_i$ is suppressed
by an additional damping factor. 

Finally, let us address the experimental implications of our result. PC's
with amplitude of about 4 nA were observed\cite{mai93} in a
micron-size GaAlAs/GaAs ring with width $w\simeq 160$ nm, and Fermi 
wavelength of electrons $\lambda_F\simeq 40$ nm, which corresponds to 
$\alpha\simeq 0.5$ and  $T_0\simeq 200$ mK. Assuming that two
concentric rings were fabricated on the same plane, the estimate of
$a_{12}$ for the above parameters and $d\sim w$ yields a small phase
shift $\delta\sim 10^{-4}$. However, this small value is due to the 
rather large distance between rings in the plane. The situation will
be significantly improved if two concentric rings were fabricated
on parallel 2D layers. Since the interlayer distance can be as
small as 3 nm\cite{mur94}, this would allow one to decrease the actual
separation between current carrying channels by 1--2 orders of
magnitude. A reliable estimate of $a_{12}$ for $d\ll w$ would require
careful treatment of the screened interaction between the two closely
spaced rings and depend on details of the structure. The
$d$--dependence (\ref{a12est}) suggests, however, that in this case  
one might expect a phase shift of several percent, which would
be within resolution of experimental observations.

\acknowledgments
The authors thank M. E. Raikh for helpful discussions. This work was
supported in part by US Department of Energy Grant No. DE-FG02-91ER45334.

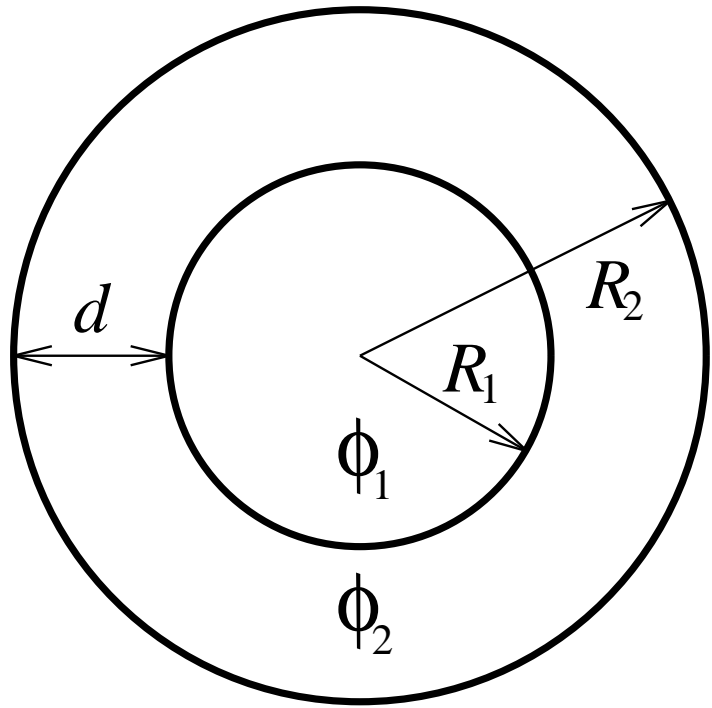
\begin{figure}
\caption{Schematic picture of two concentrically placed rings.}
\end{figure}
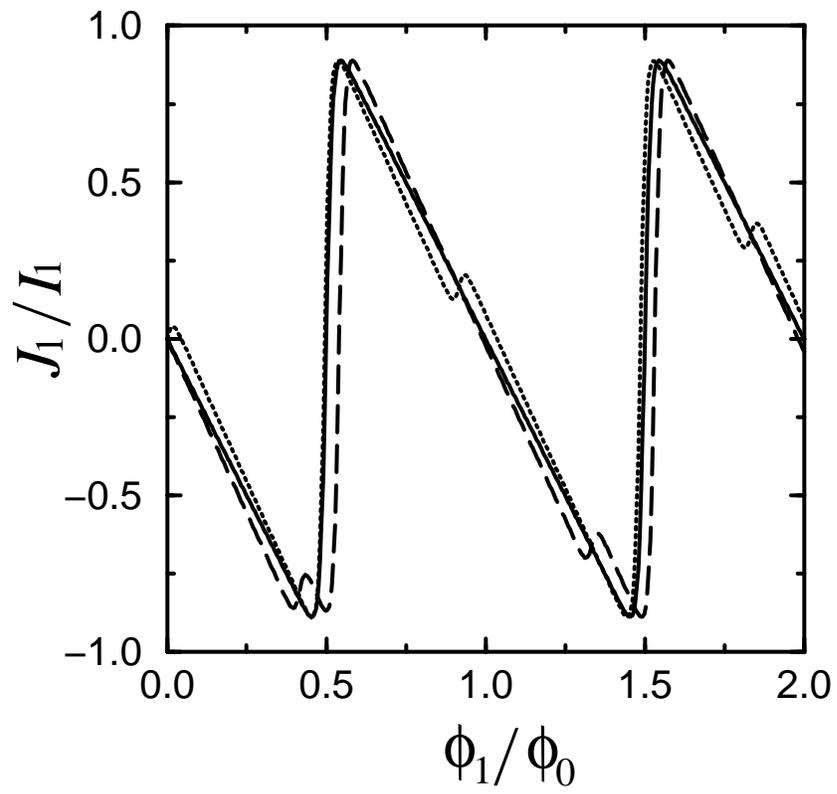
\begin{figure}
\caption{Example of PC in ring 1 at $\phi_2/\phi_1=1.1$ and $T_i/T=5.0$ for
different inter--ring couplings: 
$c_{12}/c_{ii}=0$ (solid line), and $c_{12}/c_{ii}=0.1$ with the
same (dashed line) and opposite (dotted line) parities of electron
numbers.}
\end{figure}


\begin{references}
\bibitem{remark1}We omit $x$-independent terms which ensure
proper commutation relations between fermions in different rings.
\bibitem{remark2}The ambiguity related to the ordering of non-commuting
operators in $H_{nl}$ is avoided by following the standard rule that in
all intermediate formulas the integrals over $P_i$ are performed
first.
\bibitem{remark4}A similar correction to the ground state
energy of two infinite wires was obtained in\cite{rojo92}.
\end{references}
\end{document}